\documentclass[
reprint,
superscriptaddress,
amsmath,amssymb,
prx,
floatfix,
longbibliography
]{revtex4-2}

\usepackage{graphicx}
\usepackage{bm}
\usepackage[T1]{fontenc}
\usepackage{color,soul, xcolor}
\usepackage{float}
\usepackage{natbib}
\usepackage{setspace}

\newcommand{\minus}{\scalebox{0.7}[1.0]{$-$}}
\newcommand{\plus}{\scalebox{0.7}[0.7]{$+$}}

\begin{document}

\preprint{APS/123-QED}

\title{Nanophotonic oscillators for laser conversion beyond an octave}

\author{Grant M. Brodnik}
    \email[Correspondence email address: ]{grant.brodnik@colorado.edu}
     \affiliation{Time and Frequency Division, National Institute of Standards and Technology, Boulder, CO, USA}
     \affiliation{Department of Physics, University of Colorado, Boulder, CO, USA}
\author{Haixin Liu}
     \affiliation{Time and Frequency Division, National Institute of Standards and Technology, Boulder, CO, USA}
     \affiliation{Department of Physics, University of Colorado, Boulder, CO, USA}
\author{David R. Carlson}
     \affiliation{Time and Frequency Division, National Institute of Standards and Technology, Boulder, CO, USA}
     \affiliation{Octave Photonics, Louisville, CO, USA}
\author{Jennifer A. Black}
     \affiliation{Time and Frequency Division, National Institute of Standards and Technology, Boulder, CO, USA}
\author{Scott B. Papp}
     \affiliation{Time and Frequency Division, National Institute of Standards and Technology, Boulder, CO, USA}
     \affiliation{Department of Physics, University of Colorado, Boulder, CO, USA}

\date{\today}

\begin{abstract} 

Many uses of lasers place the highest importance on access to specific wavelength bands. For example, mobilizing optical-atomic clocks for a leap in sensing requires compact lasers at frequencies spread across the visible and near infrared. Integrated photonics enables high-performance, scalable laser platforms, however, customizing laser-gain media to support wholly new bands is challenging and often prohibitively mismatched in scalability to early quantum-based sensing and information systems. Here, we demonstrate a microresonator optical-parametric oscillator (OPO) that converts a pump laser to an output wave within a frequency span exceeding an octave. We achieve phase matching for oscillation by nanopatterning the microresonator to open a photonic-crystal bandgap on the mode of the pump laser. By adjusting the nanophotonic pattern and hence the bandgap, the ratio of output OPO wave frequency span to required pump laser tuning is more than 10,000. We also demonstrate tuning the oscillator in free-spectral-range steps, more finely with temperature, and minimal additive frequency noise of the laser-conversion process. Our work shows that nanophotonics offers control of laser conversion in microresonators, bridging phase-matching of nonlinear optics and application requirements for laser designs.

\end{abstract}

\maketitle
\noindent
Sophisticated physical systems with such disparate aims as characterization of biological samples \cite{ozcelik2017scalable}, operation of quantum information protocols \cite{pino2021demonstration}, standards and sensors for time based on optical clocks \cite{mcgrew2019towards}, and precision metrology with optical sources \cite{bothwell2022resolving} demand versatility in lasers. 
Especially in the visible to near infrared, laser-wavelength access is challenging due to the limited availability and narrow operating ranges of laser-gain materials. The challenge is solved today by large, tabletop lasers based on solid-state materials and bulk nonlinear optics that provide exceptional flexibility in wavelength range, output power, and temporal profile but with substantial expense and complicated operation that hinders usability in applications.

Semiconductor lasers are less expensive and better engineered to support applications. However, the recurring effort to customize semiconductor gain materials is a barrier in expanding their use.  Moreover, achieving high performance in semiconductor lasers requires low-loss components that are often incompatible with high gain. A theme in recent laser development has been the combination of gain and photonics materials. Heterogeneous integration of gain and photonics materials on a common substrate or hybrid integration of independent laser chips and photonics chips can both overcome the challenge of semiconductor material loss, but such laser platforms currently support only select wavelength bands. To date, heterogeneous integration approaches have demonstrated
fabrication of lasers on photonic platforms in telecommunications bands \cite{huang_high-power_2019, guo_e-band_2023} and more recently below 1 $\mu$m \cite{tran_extending_2022,zhang_photonic_2023}. Still, a universal laser platform to simultaneously support visible and near-infrared wavelengths remains elusive. Hybrid integration of semiconductor lasers with photonic waveguides has demonstrated co-integration of visible sources with photonics \cite{corato-zanarella_widely_2023}, but this approach requires custom packaging and active alignment of gain material with waveguides. 

To open up laser-wavelength access in challenging spectral bands, we propose an approach bringing together semiconductor laser integration and chip-based nonlinear laser conversion. Thereby, we leverage advancing laser integration in select wavelength bands, and we develop nonlinear wavelength converters with integrated photonics that is directly compatible with laser integration. This alleviates recursive laser-gain development and provides integration. Nonlinear phase matching in integrated photonics has been extensively studied, however now we include the constraint of compatibility with laser integration. For example, both $\chi^{(2)}$ and $\chi^{(3)}$ nonlinear optical processes have enabled laser conversion with integrated photonics, including via harmonic generation
\cite{carmon2007visible,bruch2019chip,chang2016thin,nitiss2022optically,hickstein2019self,mckenna2022ultra}, difference frequency generation \cite{sahin2021difference}, four-wave mixing \cite{foster2006broad,lu2019efficient,ye2021overcoming,riemensberger2022photonic}, and optical-parametric oscillation (OPO) \cite{sayson_octave-spanning_2019,ledezma2023octave}. 
Of these processes, $\chi^{(3)}$ microresonator-based degenerate OPO is particularly appealing because of operation with a single input laser, generation of OPO output waves across wide spectral ranges, and implementation with commonly used integrated photonics.

\begin{figure*}[t] \centering \includegraphics[width=0.85\textwidth]{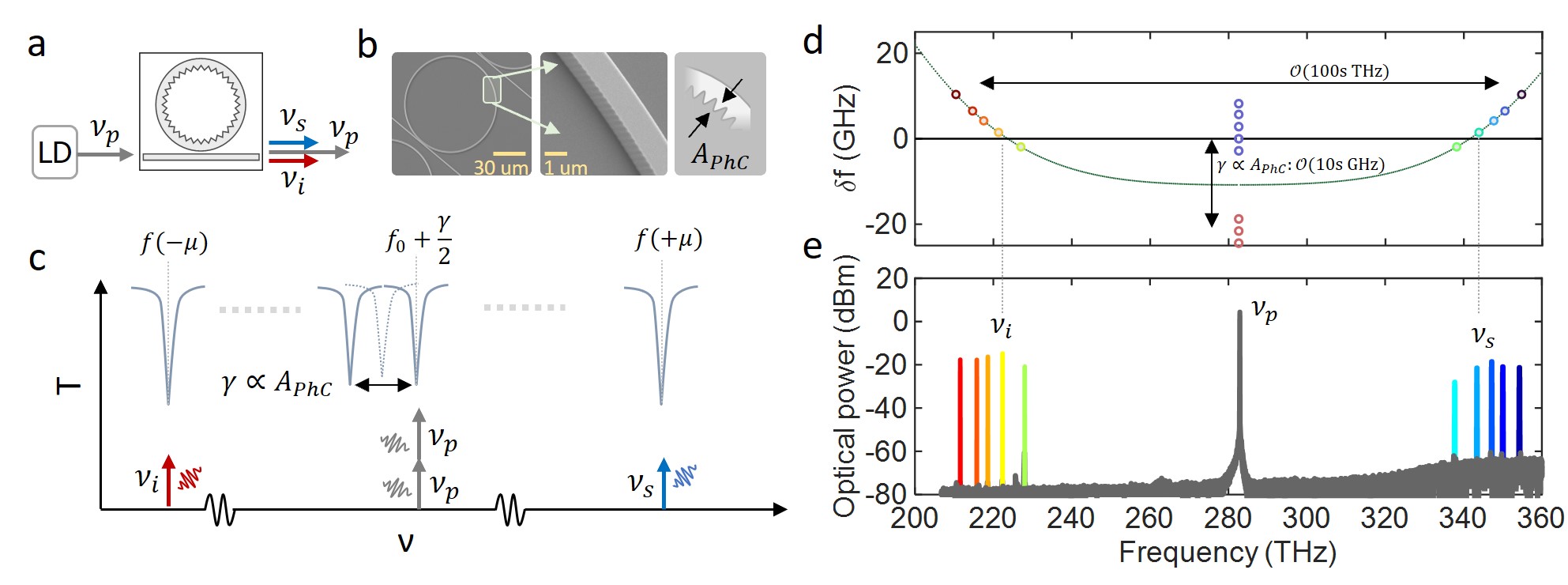}
\caption{(a) A
laser diode (LD) pumps a nanophotonic oscillator with frequency $\nu_\textrm{p}$ for OPO laser conversion
to output waves with frequency $\nu_\textrm{s}$ and $\nu_\textrm{i}$. 
(b) Scanning electron microscope (SEM) images of a nanophotonic oscillator highlighting the nanophotonic modulation with amplitude 
$\textit{A}_\textrm{PhC}$. (c) Graphical representation of phase matching which governs resonant OPO processes in a nanophotonic oscillator. (d) Frequency error,
${\delta}f$, of Eqn. 1. As we increase $\gamma$, different resonant modes $f(\pm\mu)$ are brought into phase matching for OPO. (e) Corresponding optical spectra of nanophotonic oscillation with increasing $\gamma$.}
\label{fig1}
\end{figure*}

We design microresonator OPOs by group-velocity dispersion (GVD) engineering with geometry, tailoring microresonator waveguide thickness, radius, and width \cite{lu2019milliwatt,lu_visible_2020}. Microresonator OPOs based on GVD engineering provide phase matching that conserves energy of the pump, signal, and idler waves. Moreover, such OPOs depend upon high quality factor of the microresonator modes for these waves \cite{stone2022conversion,liu2024threshold}. 
To date, the widest signal-to-idler frequency span in microresonator OPOs have a design target of near-zero GVD between the three waves. Indeed, the current limit in OPO frequency span is the required GVD engineering and challenges in fabricating microresonators with tightly controlled geometry \cite{sayson_octave-spanning_2019, lu2019milliwatt, lu_visible_2020,domeneguetti2021parametric}. 
Moreover, in operating OPOs with visible wavelength output, the increasingly normal material GVD limits control of phase matching through geometry alone.
 
Recently, OPO designs have augmented GVD engineering by geometry with the inclusion of a nanostructured periodic modulation inscribed on the inner wall of a microresonator. These so called photonic-crystal ring resonators (PhCR) induce coherent backscattering at an optical frequency determined by the nanostructure periodicity \cite{lu2014selective, yu2021spontaneous}.
This opens an optical bandgap, which has been used to control OPO phase matching  with bandgaps at the pump \cite{black_optical-parametric_2022,lu2022kerr}, at the signal wave \cite{stone_wavelength-accurate_2023}, and in the bandgap-detuned regime \cite{jin_bandgap-detuned_2024}. 
Importantly, nanostructures and geometry are two independent and direct controls of OPO phase matching.
However, the OPO signal-to-idler wave frequency span of such nanophotonic OPO has not been explored, nor has operation in the important near-zero GVD regime.

Here, we report nanophotonic oscillators that convert a pump laser to an output wave whose frequency is controllable within a span exceeding an octave. 
We use geometric GVD engineering to access the near-zero GVD regime between the OPO waves, and we use a photonic-crystal bandgap to provide phase matching for broadband OPO. 
This enables robust laser conversion and access to an unprecedented OPO wave frequency span in nanostructured microresonators. 
Importantly, our approach simplifies the required pump laser tuning range for a microresonator OPO; in our experiments the pump tuning is 10,000 times less than the OPO wave frequency span. 
We realize nanophotonic oscillators on the tantalum pentoxide (Ta\textsubscript{2}O\textsubscript{5}), or tantala, platform, which has high Kerr nonlinearity, low stress, and high quality factor across the visible and near-infrared wavelengths \cite{jung_tantala_2021}. Moreover, tantala supports GVD engineering \cite{black_group-velocity-dispersion_2021} and heterogeneous integration \cite{jafari2022heterogeneous,dorche2023heterogeneously,blumenthal2020photonic}.
Experimentally, we demonstrate nanophotonic oscillators, which take a pump laser near 1062 nm and create output OPO signal and idler waves. We demonstrate a maximum OPO wave frequency span from 749 nm to 1806 nm that corresponds to >230 THz with pump-laser tuning <30 GHz. 
These experiments demonstrate the principle of laser-wavelength access while requiring minimal pump laser tuning, which is a critical innovation for laser-wavelength access on a chip. 
In view of interest to use the devices for applications, we characterize several aspects of the output OPO, including thermal fine-frequency tuning and additive frequency noise of the laser-conversion process.

\vspace{5mm}
\noindent\textbf{Nanophotonic wavelength converters} \newline
Nanophotonic oscillators enable laser conversion through resonant phase matching of the pump, signal, and idler waves \cite{kippenberg_kerr-nonlinearity_2004}. 
We characterize the frequency error ${\delta}f(\mu)$ for phase matching as   
\begin{equation}
{\delta}f(\mu) = 2(f_0\pm\gamma/2) - f(\plus\mu) - f(\minus\mu)
\label{df_gamma}
\end{equation}
where the resonant frequencies of a microresonator are $f(\mu)$ and $\mu = m - m_p$ is the azimuthal mode number, $m$, relative to the pump, $m_p$. The pump mode $f(\mu = 0)$ is $f_0$, which we split with a photonic-crystal bandgap of magnitude $\gamma$, thereby shifting its frequency without affecting the GVD design.
The pump mode, $f_0 \pm \gamma$/2, 
phase matches to modes symmetric about the pump, $f(\pm\mu)$, when Eqn. 1 is positive to within the pump mode Kerr shift 
\cite{black_optical-parametric_2022,stone2022conversion, stone_wavelength-accurate_2023}. 
We operate nanophotonic oscillators in the near-zero GVD regime and use $\gamma$ to control the signal and idler wave frequencies with minimal pump laser frequency tuning. 
Figure 1(a) presents the concept of nanophotonic oscillators where we convert a pump laser with frequency $\nu_\textrm{p}$ to signal and idler waves with frequency $\nu_\textrm{s}$ and $\nu_\textrm{i}$. We instill $\gamma$ by inscribing a nanopattern into the microresonator sidewall with modulation amplitude $A_\textrm{PhC}$ as seen in Fig. 1(b).Graphically, we can understand nanophotonic oscillator laser conversion through the mode structure depicted in Fig. 1(c). Operationally, we tune the pump laser to resonance $\nu_\textrm{p} = f_0 + \gamma/2$ to initiate OPO-based laser conversion to $\nu_\textrm{s}$ and $\nu_\textrm{s}$, which abides strict energy conservation $2\nu_\textrm{p} = \nu_\textrm{s} + \nu_\textrm{i}$ \cite{black2023nonlinear}.

We realize this method in practice, demonstrating $\gamma$-controlled laser conversion. 
To implement nanophotonic oscillators, we use PhCRs, which provide a broadband and high-finesse microresonator with programmable $\gamma$ to control OPO phase matching.
Figure 1(d) plots ${\delta}f$ of such a nanophotonic oscillator where we see that varying $\gamma$ enables alignment of $f_0 + \gamma/2$ to different pairs of $f(\pm\mu)$, thereby modifying OPO phase matching to control $\nu_\textrm{s}$ and $\nu_\textrm{i}$. 
We transfer the ${\delta}f$ design of Fig. 1(d) to devices in the tantala platform (see Methods), demonstrating laser conversion as seen in Fig. 1(e) in which we plot optical spectra for several $\gamma$ settings with otherwise identical GVD. 
The pump $\nu_\textrm{p}$ needs tune only several GHz to track $\gamma$/2 while $\nu_\textrm{s}$ and $\nu_\textrm{i}$ are tuned across $\approx$ 20 THz with an OPO wave frequency span exceeding 100 THz. 

\vspace{5mm}
\noindent\textbf{Device characteristics} \newline
\noindent 
In addition to satisfying Eqn. 1 for OPO phase matching, we require a high quality factor to realize nanophotonic oscillators \cite{lu2019milliwatt, stone2022conversion, stone2022efficient}. 
To characterize the quality factor of devices, we measure transmission spectra of three widely tunable external cavity lasers near the 980 nm, 1064 nm, and 1300 nm wavelength bands.
Figure 2(a) presents the loaded, intrinsic, and coupling quality factors ($Q_\textrm{L}$, $Q_\textrm{i}$, and $Q_\textrm{c}$) as a function of frequency for a nanophotonic oscillator device.
We measure $Q_\textrm{i}$ $\approx$ 0.9 x 10$^6$ across the three wavelength bands, suitable for efficient OPO. 
Moreover, we measure the wavelength-dependent $Q_\textrm{c}$ to agree with three-dimensional, finite-difference time-domain modelling
within fabrication tolerance (Methods) \cite{moille2019broadband}. 
\begin{figure*}[ht] \centering \includegraphics[width=0.85\textwidth]{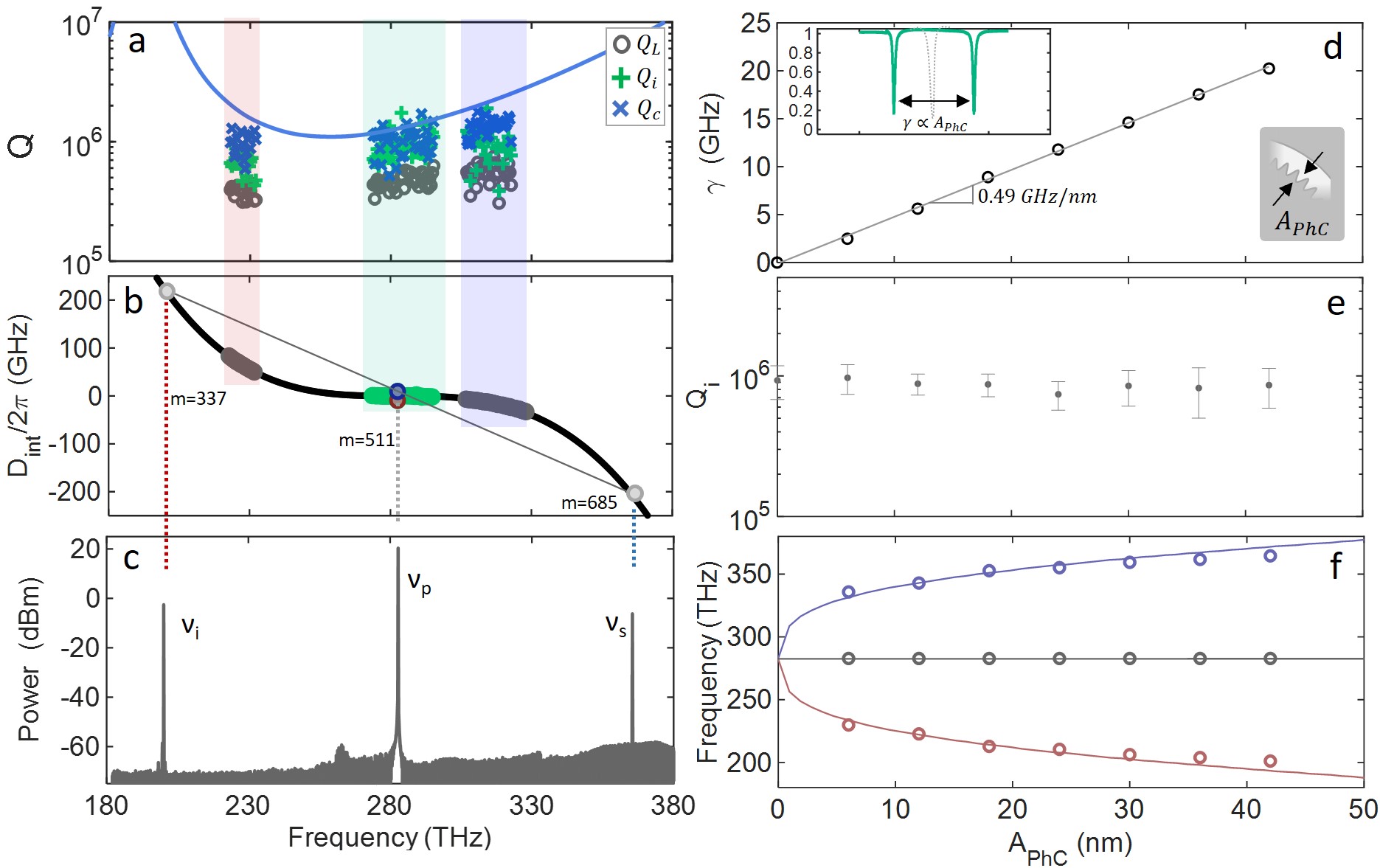}
\caption{(a) Measured $Q_\textrm{L}$, $Q_\textrm{i}$, and $Q_\textrm{c}$ (colored markers) across three spectral regions and modelled $Q_\textrm{c}$ (blue line). (b) Experimental (colored circles) and modelled (black circles) $D_\textrm{int}$/2$\pi$. (c) OPO spectrum for the device characterized in (a) and (b). (d) Measured $\gamma$ vs. $A_\textrm{PhC}$. Left inset shows example pump resonance  $f_0\pm\gamma/2$ and right inset shows $A_\textrm{PhC}$ inner sidewall modulation. (e) $Q_\textrm{i}$ vs. $A_\textrm{PhC}$. (f) Measured (circles) and modelled (lines) OPO frequencies vs. $A_\textrm{PhC}$, highlighting nanophotonic control of OPO.}
\label{fig2}
\end{figure*}

To explore the accuracy of $\gamma$-controlled phase matching in our nanophotonic oscillators, we measure the GVD and pump the devices above the OPO threshold power to validate Eqn. 1.
We express the resonant angular frequencies, $\omega$, of a microresonator as a function of $\mu$ using a Taylor expansion: 

\begin{equation} \label{dint_polynomial}
\begin{split}
 \omega(\mu) & = 2\pi f({\mu}) = {2\pi f_0} + {D_1}\mu + \frac{D_2\mu^2}{2} + \frac{D_3\mu^3}{6} + ... \\
 & = {\omega_0} + {D_1}\mu + D_\textrm{int}.
\end{split}
\end{equation}
The integrated dispersion, $D_\textrm{int}$, relates the change in resonant angular frequencies relative to the free spectral range (FSR $= D_1/2\pi$) at the pump \cite{fujii2020dispersion}. 
From our measured transmission spectra, we can directly determine $D_\textrm{int}$ and compare with finite element method simulations of the microresonator eigenfrequencies (see Methods for more). 
Figure 2(b) presents measured (colored circles) and modelled (black circles) $D_\textrm{int}/2\pi$ as a function of frequency. 
We find the nanophotonic oscillator pump resonance $f_0+\gamma/2$ with $m_\textrm{p}$ = 511 satisfies Eqn. 1 at modes $m$ = 685 and $m$ = 337 ($\mu = \pm$ 174). 
In terms of $D_\textrm{int}$, OPO phase matching occurs when we can draw a straight line through $f_0+\gamma/2$ and $f(\pm{\mu})$ as seen in Fig. 2(b) \cite{black2023nonlinear}.
We validate this by pumping the nanophotonic oscillator with $D_\textrm{int}$ from Fig. 2(b) and recording the output OPO spectrum as seen in Fig. 2(c). We find that when pumping $f_0+\gamma/2$ at $\nu_\textrm{p}$ = 282.8 THz, the nanophotonic oscillator provides laser conversion to output waves with frequencies $\nu_\textrm{s}$ = 365.5 THz and $\nu_\textrm{i}$ = 200.1 THz.

Deterministic control of $\gamma$ is of critical importance for laser conversion with nanophotonic oscillators. 
Therefore, we characterize $\gamma$ as a function of $A_\textrm{PhC}$ and find a linear response with slope $\approx$ 0.5 GHz/nm up to $A_\textrm{PhC}$ = 50 nm as seen in Fig. 2(d). 
Remarkably, despite the presence of the nanostructure, we measure little $Q_\textrm{i}$ variation or degradation over the full range of $\gamma$ < 25 GHz as seen in Fig. 2(e). We control the laser conversion process by use of $A_\textrm{PhC}$ as shown in Fig. 2(f).
Here, we find that $A_\textrm{PhC}$ < 50 nm results in a large change of OPO phase matching governed by Eqn. 1, with control of both $\nu_\textrm{s}$ and $\nu_\textrm{i}$ across > 50 THz and maximum OPO wave frequency span exceeding 150 THz of optical bandwidth. 

\vspace{5mm}

\begin{figure*}[ht] \centering \includegraphics[width=0.85\textwidth]{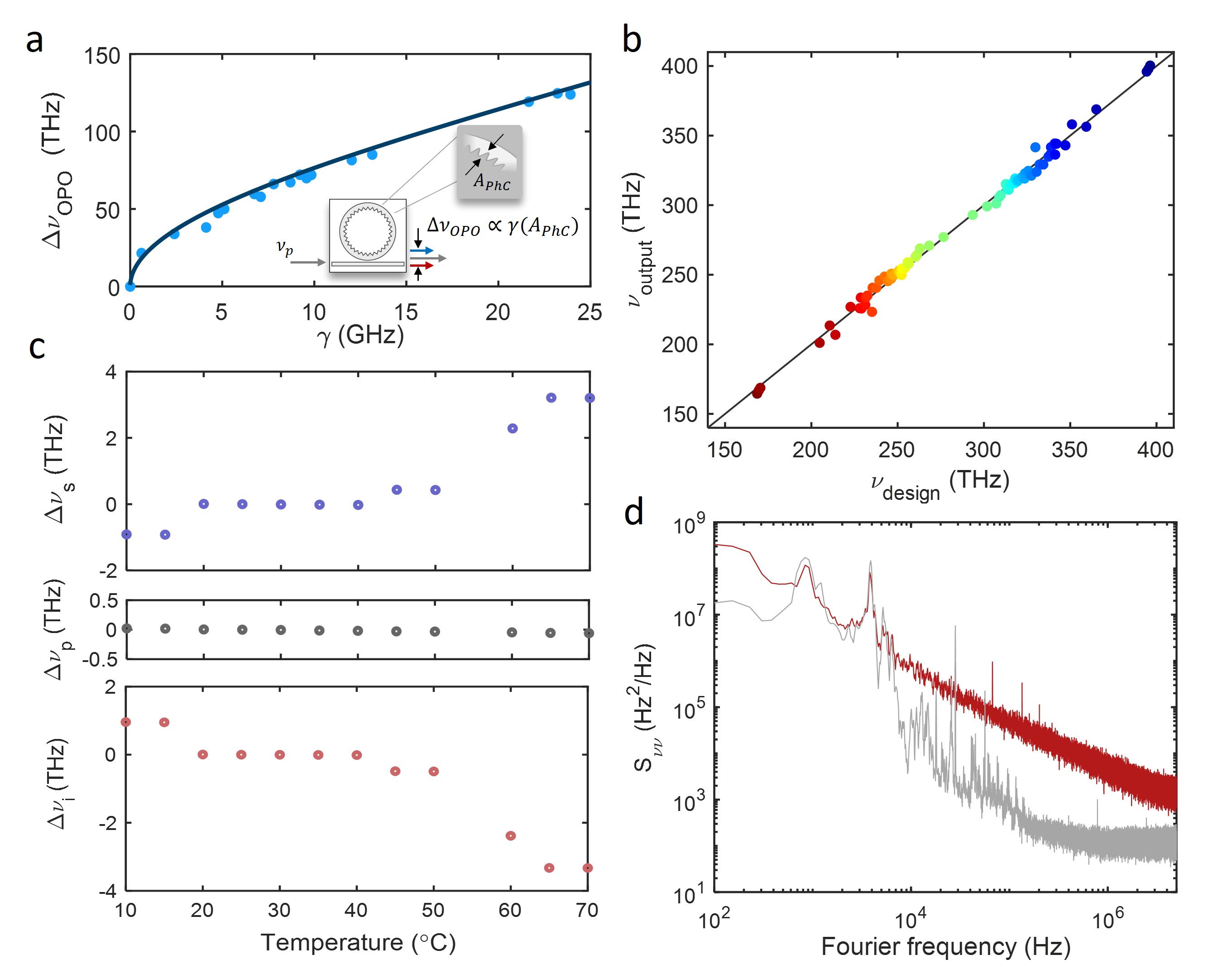}
\caption{ (a) Measured (points) and modelled (line) $\Delta\nu_{OPO}$ as a function of $\gamma$ for an otherwise fixed GVD. (b) Measured (points) and modelled (line) $\nu_\textrm{output}$ vs $\nu_\textrm{design}$. (c) Thermal response of OPO wave frequencies in a single nanophotonic oscillator. (d) Measured frequency noise for an OPO with $\nu_\textrm{p}$ = 282.8 THz (grey) and 
$\nu_\textrm{i}$ = 219.3 THz (red).}
\label{fig3}
\end{figure*}

\noindent \textbf{Frequency design and control} \\
\noindent 
Leveraging nanophotonic oscillators, we demonstrate wavelength access beyond an octave through $\gamma$-controlled laser conversion. We measure OPO spectra of devices pumped above threshold, and 
we extract $\nu_\textrm{s}$ and $\nu_\textrm{i}$ to determine the output OPO signal-to-idler frequency span, which we denote $\Delta\nu_\textrm{OPO}$ = $\nu_\textrm{s}$ - $\nu_\textrm{i}$.
Figure 3(a) shows measurements (points) and modelling (line) of $\Delta\nu_\textrm{OPO}$ as a function of $\gamma$, predicted by the phase-matching condition of Eqn. 1.  
Figure 3(b) presents a compilation of measured output OPO wave frequencies, $\nu_\textrm{s}, \nu_\textrm{i}$, which we denote as the $\nu_\textrm{output}$ (points) for various nanophotonic oscillators and compare with the Eqn. 1 prediction, which we denote as $\nu_\textrm{design}$ (line) of the OPO; see Methods. These data assess the maximum wave span of the OPO.

Provided the wavelength access of nanophotonic oscillators, we characterize other important operational details for applications, including thermal frequency tuning and power-spectral noise density of output waves. Figure 3(c) shows the variation of OPO wave frequencies in a single device as we vary the temperature from 10$^\circ$C to 70$^\circ$ by use of a thermoelectric cooler.
At room temperature, T = 20$^\circ$C, $\nu_\textrm{s}$ = 365.5 THz and $\nu_\textrm{i}$ = 200.1 THz ($\Delta\nu_\textrm{OPO}$ = 165.4 THz). We then measure the three OPO waves as a function of temperature $\nu_\textrm{s,p,i}(T)$, and plot the variation from the room temperature measurement, $\Delta\nu_\textrm{s,p,i}$ = $|\nu_\textrm{s,p,i}$(20$^\circ$C) - $\nu_{s,p,i}(T)|$. During thermal tuning, we shift $\nu_\textrm{p}$ to track the thermo-optic shift of the pump resonance mode with slope $\approx$ \minus1.4 GHz/C, consistent with tantala's reported thermo-optic coefficient \cite{jung_tantala_2021}. However, across the measured temperature range, $\nu_\textrm{s}$ and $\nu_\textrm{i}$ tune by $\approx$ 4 THz discontinuously, which we attribute to thermally sensitive phase matching \cite{lu_visible_2020}. We also measure the optical frequency noise power spectral density, $S_\nu(f)$, for $\nu_\textrm{p}$ and $\nu_\textrm{i}$ (Fig. 3(d)), calculating the integral linewidth by use of the 1/$\pi$ and $\beta$-separation techniques \cite{ liang2015ultralow,di2010simple}. We find the pump laser ($\nu_\textrm{p}$ = 282.8 THz) to have a 1/$\pi$-integrated linewidth of 28 $\pm$ 3 kHz and a $\beta$-separation linewidth of 862 $\pm$ 61 kHz integrated down to Fourier frequency of 78 Hz. 
The corresponding free-running $\nu_\textrm{i}$ = 219.3 THz has a 1/$\pi$-integrated linewidth of 89 $\pm$ 6 kHz and a $\beta$-separation linewidth of 1.03 $\pm$ 0.07 MHz. 
We attribute the additional optical-frequency noise of $\nu_\textrm{i}$ to thermo-refractive noise due to the characteristic bandwidth of the power-spectral density \cite{drake2020thermal}.

\vspace{5mm}
\noindent\textbf{Discussion} \newline
\noindent
In summary, we have demonstrated a nanophotonic approach to achieve phase matching for broadband OPO laser conversion into the visible and near-infrared. Moreover, thermal fine-frequency tuning and the laser conversion process contribute minimal additive frequency noise in nanophotonic oscillators. 
By leveraging the precise phase-matching control afforded by nanophotonic oscillators, we realize laser conversion spanning more than an octave with pump tunability of only tens of gigahertz, a ratio exceeding 10,000. Nanophotonic oscillators enable the broadband, configurable gain afforded by nonlinear optics to operate in concert with existing integrated pump lasers, avoiding the necessity to re-customize laser gain materials in underdeveloped spectral regions. Thereby, nanophotonic oscillators address the challenge of laser-wavelength access, in particular, reaching across the short-wave infrared and into the visible.

\vspace{5mm}
\noindent\textbf{Methods} \newline
\noindent
\textbf{Nanophotonic oscillator design} \newline
We design nanophotonic oscillator GVD to support OPO by satisfying Eqn. 1. 
To determine the GVD of a nanophotonic oscaillator, we first solve for the ordinary microresonator ($A_\textrm{PhC} = 0$) eigenfrequencies, using a finite element method solver.
We include the bulk material GVD of the bottom cladding (silica)
and the waveguide layer (tantala \cite{black_group-velocity-dispersion_2021}) and the waveguide geometry, including microresonator radius, waveguide width, and thickness. 
To target specific OPO output wave frequencies, we use Eqn. 1 to determine the necessary $\gamma$ for a given ordinary microresonator GVD. 
We then map the required $A_\textrm{PhC}$ to achieve target $\gamma$ using the results from Fig. 2(d). 
We modelled various geometries of tantala waveguides with a measured thickness 566 $\pm$ 5 nm and an under-etched tantala pedestal height of 21 $\pm$ 2 nm, within our fabrication variation. As seen in the figures, these geometric ranges provide good agreement with the measurements. 
Finally, to efficiently couple light into the nanophotonic oscillator, we employ a broadband pulley coupler with a pulley-coupler length of 5 $\mu$m, a bus waveguide width of 650 nm, and a bus-to-resonator gap of 180 nm \cite{moille2019broadband}. 
We employ commercially available software to model the anticipated $Q_\textrm{c}$ using the three-dimensional finite-difference time-domain method. We find the calculated $Q_\textrm{c}$ to be slightly larger than the measurement (see Fig. 2(a)), suggesting the devices are more over-coupled than the design, which is in good agreement with our GVD modelling of an under-etched tantala film. 

\vspace{3mm}
\noindent\textbf{Experimental testing}
\newline
Measuring passive characteristics of nanophotonic oscillators requires precise frequency tracking of our tunable lasers.
We employ fibre Mach-Zehnder interferometers (MZIs) with RF-calibrated free spectral ranges (FSR) as optical frequency rulers when measuring quality factors, $D_{\textrm{int}}$, and $\gamma$ vs. $A_{PhC}$ (see Fig. 2a, 2b, and 2d). We split the output of our tunable lasers into two paths, with one path coupled to the PhCR under test and the second path passing through the fibre MZI. Simultaneously recording both the MZI and PhCR transmission while frequency sweeping the tunable laser enables direct, continuous mapping of relative optical frequency by unwrapping the sinusoidal output of the MZI. We include fibre dispersion in the FSR calibration, which is particularly important when measuring GVD. We use three independent, widely tunable lasers to measure $D_\textrm{int}$, with their corresponding measurements stitched together for the plot in Fig. 2b.
Additionally, we use an RF-calibrated MZI as an optical frequency-to-voltage discriminator for the frequency noise measurements presented in Fig. 3d. For the OPO frequency noise measurement, the idler is optically filtered using a free space bandpass filter.

\vspace{3mm}
\noindent\textbf{Fabrication procedure} \newline
Our fabrication process begins with an oxidized silicon wafer with an ion-beam-sputtered tantala waveguide layer (FiveNine Optics) with target thickness 570 nm. 
We transfer our designs to the waveguide layer using electron-beam lithography and a fluorine inductively coupled plasma-reactive ion etch. 
The top air clad devices are singulated into chips using UV lithography and a reactive ion dry etch. Finally, we use an overnight thermal anneal in air at 500$^{\circ}$C to reduce material defects. This 3-inch, wafer-scale process affords more than 3000 independent PhCR laser converters per fabrication cycle. 

\vspace{5mm}

\noindent\textbf{Acknowledgments} \newline
\noindent
We thank Lindell Williams and Yan Jin for reviewing the manuscript. This research has been funded by the DARPA LUMOS program HR0011-20-2-0046, AFOSR FA9550-20-1-0004 Project Number 19RT1019, NSF Quantum Leap Challenge Institute Award OMA - 2016244, and NIST.  This work is a contribution of the US Government and is not subject to US copyright. Mention of specific companies or trade names is for scientific communication only and does not constitute an endorsement by NIST. 

\clearpage

\bibliography{Brodnik_Nanophotonic_oscillators_MAIN} 

\end{document}